\documentclass{jssc}

\def\textsubscript#1%
{$_{\text{#1}}$}

\def\cdd{\mbox{\boldmath$\cdot$}~}

\input{amssym.def}
\theoremstyle{remark}
\usepackage{algorithm,algorithmic}

\newcommand{\onen}{\frac{1}{n}}

\newcommand{\sumn}{\sum_{i=1}^{n}}
\newcommand{\oner}{\frac{1}{r}}%
\newcommand{\sumr}{\sum_{i=1}^{r}}%

\newcommand{\Exp}{\mathbb{E}}

\newcommand{\Var}{\mathbb{V}}

\newcommand{\opt}{\mathrm{opt}}
\newcommand{\bpi}{\bm{\eta}}

\makeatletter
\def\@oddfoot{\hfill}
\newcount\shumeicount
\def\setshumei#1#2#3{%
  \shumeicount=\count0
  \def\@oddhead{%
    \raise-5pt\hbox to0pt{\vrule width\hsize height 0pt depth 0.4pt\hss}\relax
    \ifnum \shumeicount=\count0
      \raise-7pt\hbox to0pt{\vrule width\hsize height 0pt depth 0.4pt\hss}\relax
      #1
    \else
      \ifodd\count0
        #2
      \else
        #3
       \fi
     \fi
  }%
}
\makeatother
\makeatletter
\def\@oddfoot{\hfill}
\newcount\shujiaocount
\def\setshujiao{%
  \shujiaocount=\count0
  \def\@oddfoot{%
      \ifodd\count0
      \else
      \fi
  }%
}
\makeatother
\def\biaoti#1#2#3#4{{
  \vspace*{0.3cm}
  \begin{flushleft} \Large\bf #1\end{flushleft}
  \vspace*{-0.2cm}
      \begin{flushleft}
      \bf #2
      \end{flushleft}
      \footnotetext{\hspace{-6mm} #3\\ #4}}}

\def\dshm#1#2#3#4
{\setshumei{J Syst Sci Complex (#1) #2:
{\thepage--\pageref{LastPage}}\hfill}
            {\hfill {\small #3}\hfill\hbox to0pt{\hss\thepage}}
            {\hbox to0pt{\thepage\hss}\hfill {\small #4}\hfill
            }
            \setshujiao}
\def\drd#1#2
{{\vskip 1cm\small \begin{flushleft}
 #1 \\
 #2\\
\copyright The Editorial Office of JSSC \&  Springer-Verlag Berlin
Heidelberg 2014
\end{flushleft}}}

\def\hat{\widehat}

\def\bar{\overline}
\def\epsilon{\varepsilon}

\begin{document}

\biaoti{Most Likely Optimal Subsampled Markov Chain Monte Carlo}%
{Guanyu \uppercase{Hu}$^{1,2}$  %
  \cdd 
HaiYing \uppercase{Wang}$^1$ %
}%
{Address: %
215 Glenbrook Rd. U-4120
Storrs, CT 06269-4120\\
University of Connecticut$^{1}$; University of Missouri-Columbia$^{2}$\\
    Email: guanyu.hu@uconn.edu; haiying.wang@uconn.edu} %
{$^*$This research was supported by U.S. National Science Fundation under Grant No 1812013.\\
{$^\diamond${\it This paper was recommended for publication by
Editor . }}}

\drd{DOI: }{Received: x x 20xx}{ / Revised: x x 20xx}

\dshm{20XX}{XX}{A TEMPLATE FOR JOURNAL}{\uppercase{Hu Guanyu} $\cdd$ \uppercase{Wang HaiYing} $\cdd$
}

\Abstract{Markov Chain Monte Carlo (MCMC) requires to evaluate the full data likelihood at different parameter values iteratively and is often computationally infeasible for large data sets.  In this paper, we propose to approximate the log-likelihood with subsamples taken according to nonuniform subsampling probabilities, and derive the most likely optimal (MLO) subsampling probabilities for better approximation. Compared with existing subsampled MCMC algorithm with equal subsampling probabilities, our MLO subsampled MCMC has a higher estimation efficiency with the same subsampling ratio.  We also derive a formula using the asymptotic distribution of the subsampled log-likelihood to determine the required subsample size in each MCMC iteration for a given level of precision. This formula is used to develop an adaptive version of the MLO subsampled MCMC algorithm. Numerical experiments demonstrate that the proposed method outperforms the uniform subsampled MCMC.}      %


\section{Introduction}
\label{sec:introduction}
Bayesian methods became popular since 1990s due to the advance of
computing technology and the introduction of powerful sampling
algorithm like Markov Chain Monte Carlo (MCMC). However, posterior
sampling through MCMC is computationally demanding, especially with
large data sets. When a data set has a large number of observations, the
MCMC method may take a long time to run because it requires to
evaluate the likelihood function in each iteration on the full
data. There are two major approaches to speed up MCMC algorithms. The
first approach utilizes parallel computing; it partitions the data
into small pieces and computes sub-posteriors for each piece in
parallel, see \cite{wang2013parallelizing,scott2016bayes}. The other
approach is to use a subsample of the data in each MCMC iteration to
speed up the algorithm, e.g., 
\cite{bardenet2014towards,korattikara2014austerity}. This paper is
about the subsampling approach.

A standard way to conduct MCMC is to apply the Metropolis-Hastings
(MH) algorithm \cite{gelman2014bayesian,robert2004monte}, and we will
focus on this algorithm. The MH algorithm requires to evaluate the
full data likelihood at two different values of the parameter in each iteration. We
propose to approximate the full data log-likelihood with weighted log-likelihood calculated from subsamples taken according to nonuniform
subsampling probabilities. The subsampled log-likelihood estimator is unbiased approximations of the full data log-likelihood. We then derive the
most likely optimal subsampling probabilities to better approximate the full data log-likelihood. Compared with existing uniform subsampled MH
algorithm \citep{bardenet2014towards,korattikara2014austerity}, our
algorithm has a higher estimation efficiency with the same subsampling
ratio or requires a smaller subsample size for the same level of
approximation precision. Another contribution of this paper is that we
introduce a subsample size calculation formula to determine required
subsample size adaptively for a given precision in each MH iteration. Unlike the sample size determination rule used in
\cite{bardenet2014towards}, which relies on an upper bound of the
approximation error, our formula is based on the asymptotic
distribution of the approximation error. As a result, the required
subsample sizes are typically smaller than that required by \cite{bardenet2014towards}.

The rest of the paper is organized as follows. Section~\ref{sec:backgr-relat-work} introduces notations for the problem of interest, the traditional MH algorithm, and the existing uniform subsampled MH algorithm. Section~\ref{sec:weight-subs-mcmc} presents the MLO subsampling probabilities and resultant MH algorithms, one with a fixed subsample size and the other with adaptive subsample sizes. Section~\ref{sec:simulation} uses numerical experiments to evaluate our method and draws comparisons with uniform subsampled MCMC algorithm. Section~\ref{sec:covertype-data} illustrates the proposed methods on a real data set. Section~\ref{sec:conclusion} concludes with a brief summary of the paper and possible future research topics.

\section{Background and related work}
\label{sec:backgr-relat-work}
Consider a data set with $n$ data points, $X=\{x_{1},...,x_{n}\}$, for which the underlying distribution depends on a $p$ dimensional parameter vector $\theta$. Given a value of the parameter $\theta$, we assume that the data are conditionally independent with associated likelihood $p(X|\theta)=\prod_{i=1}^{n}p(x_{i}\vert\theta)$. 
In Bayesian approach, $\theta$ is assume to be random with a prior distribution, say $p(\theta)$. Bayesian inference relies on the posterior distribution of $\theta$, the conditional distribution of $\theta$ given the data,\begin{equation}\label{eq:5}
  \pi(\theta)=\frac{p(X|\theta)\times p(\theta)}{\int p(X|\theta)\times p(\theta)d \theta}
  \propto p(\theta)\prod_{i=1}^{n}p(x_{i}\vert\theta).
\end{equation}
Statistical inference often requires to calculate a functional of $\pi(\theta)$ such as the posterior mean, $\int\theta\pi(\theta)d\theta$, which is the Bayes estimator under the squared loss. 
In most applications, $\pi(\theta)$ has a complicated expression and the functional is analytically  %
infeasible to find. For this scenario, one often has to use MCMC methods to generate samples from the posterior distribution for statistical inference.

\subsection{The standard Metropolis-Hastings algorithm}
\label{ss:MH} 
The MH algorithm is a widely used method to sample approximately from $\pi(\theta)$. %
This algorithm needs a conditional proposal distribution, say $q(.\vert\theta)$, to generate a candidate parameter value $\theta'$, and then the posterior density need to be evaluated at $\theta'$ to determine if $\theta'$ is accepted or rejected as the next step value of the algorithm. In practice, $q(.\vert\theta)$ must be a distribution from which it is easy to simulate observations. The MH algorithm produces a Markov chain with the posterior as its equilibrium distribution. Thus, after sufficient number of iterations, the MH algorithm produces observations from the posterior distribution. 
For completeness and ease of discussion, we present the standard MH algorithm in the following Algorithm~\ref{alg:1}:
\begin{algorithm}[H]
  \caption{Metropolis-Hastings algorithm}
  \label{alg:1}
  \begin{algorithmic}
    \FOR{$k\leftarrow1$ \TO $N$}
    \STATE $\theta \leftarrow \theta_{k-1}$
    \STATE $\theta'\sim q(.\vert\theta)$
    \STATE $u\sim U{(0,1)}$
    \STATE $\alpha=\frac{\pi(\theta')q(\theta|\theta')}{\pi(\theta)q(\theta'|\theta)}$
    \IF {$\alpha>u$}
    \STATE $\theta_k\leftarrow\theta'$ \{Accept\}
    \ELSE
    \STATE $\theta_k\leftarrow\theta$ \{Reject\}
    \ENDIF
    \STATE Return $\theta_k$, $k=1, ..., N$
    \ENDFOR
    \end{algorithmic}
\end{algorithm}
\subsection{Metropolis-Hastings algorithm with subsampled likelihood}
In the standard MH algorithm~\ref{alg:1}, one has to evaluate $\pi(\cdot)$ at both $\theta$ and $\theta'$ in each iteration. From \eqref{eq:5}, this requires to evaluate the full data likelihood at each iteration, which is computationally demanding for large data sets.
\cite{bardenet2014towards, korattikara2014austerity} proposed to approximate the full data log-likelihood using uniform subsamples to speed up MCMC methods. We briefly discuss the rationale of this idea here.

The accept-or-reject step in Algorithm~\ref{alg:1} is determined by the relative magnitude between $\alpha$ and $u$. Note that $\alpha>u$ if and only if 
\begin{align}\label{eq:7}
  \Lambda_n(\theta,\theta')\equiv
  \ell_n(\theta')-\ell_n(\theta)>
  \onen\log\left[
  u\frac{p(\theta)q(\theta'|\theta)}{p(\theta')q(\theta|\theta')}
  \right],
\end{align}
where
\begin{align}\label{eq:loglikelihood}
  \ell_n(\theta)=\onen\sumn\log{p(x_i|\theta)}
\end{align}
is the full data log-likelihood. In each MH iteration, the major computing burden is to calculate the full data log-likelihood at $\theta$ and $\theta'$. However, since $\ell_n(\theta)$ is in a form of average, it can be well approximated by using a subsample, and that is the basic idea of subsampled MH algorithm to speed up MCMC.

In \cite{bardenet2014towards, korattikara2014austerity}, the authors proposed to take uniform subsamples to approximate the full data log-likelihood difference $\Lambda_n(\theta,\theta')$. Let $x^*_1,\cdots,x^*_r$ be random sample of size $r$ from the full data taken without replacement according to uniform subsampling probabilities. 
Instead of computing $\Lambda_n(\theta,\theta')$ on the full data set, they proposed to approximate $\Lambda_n(\theta,\theta')$ by %
\begin{align}
  \Lambda^*_{ur}(\theta,\theta')=
  \oner\sumr\log{p(x^*_i|\theta')}-\oner\sumr\log{p(x^*_i|\theta)}.
	\label{subsample log likelihood}
\end{align}

 To determine the required subsample sizes, \cite{korattikara2014austerity} put the accept-or-reject step of MH in a framework of hypothesis testing and treated the inequality in \eqref{eq:7} as the null hypothesis. Given the full data, $\theta$, $\theta'$, and $u$, if we write $\mu=\frac{1}{n}\sum_{i=1}^n(\log p(x_i|\theta')-\log p(x_i|\theta'))$ and $\mu_0=\frac{1}{n}\log\left[
  u\frac{p(\theta)q(\theta'|\theta)}{p(\theta')q(\theta|\theta')}
  \right]$, then to determine if the inequality in \eqref{eq:7} is true or not, it is equivalent to test $H_0: \mu>\mu_0$ v.s. $H_a: \mu\le\mu_0$. 
This is just a hypothesis test for the mean, so the random subsample mean, $\Lambda^*_{ur}(\theta,\theta')$, can be used to form a test statistic. \citep{korattikara2014austerity}'s subsample size determination rule is to take enough subsample so that
  the p-value of the hypothesis test $H_0: \mu=\mu_0$ v.s. $H_a: \mu\neq\mu_0$ is smaller than a threshold. %

For a given level of precision, 
\cite{bardenet2014towards} proposed to determine the required subsamples size by using the following concentration inequality to bound the error of $\Lambda^*_r(\theta,\theta')$ in approximating $\Lambda_n(\theta,\theta')$: 
\begin{align}
	P(|\Lambda^*_r(\theta,\theta')-\Lambda_n(\theta,\theta')|\leq c_r)\geq 1-\delta_r,
	\label{subsample ineq}
\end{align}
for $\delta_r$, where $c_r=C_{\theta,\theta'}\sqrt{\frac{2(1-f^*_r)\log (2/\delta_r)}{r}}$, $C_{\theta,\theta'}=\max_{1\leq i \leq n}|\log p(x_i|\theta')-\log p(x_i|\theta)|$, and $f^*_r=\frac{r-1}{n}$ is approximately the fraction of used samples. 
Based on the inequality \eqref{subsample ineq}, they developed an adaptive procedure to determine the required subsample size in each MH iteration as
\begin{align}
	T=n\land \inf \{t \geq 1:|\Lambda^*_r(\theta,\theta')-\Lambda_n(\theta,\theta')|< c_r\},
	\label{eq:stop time}
\end{align}
where $a\land b$ denotes the minimum of $a$ and $b$. Using this adaptive subsample size \eqref{eq:stop time} and subsampled log-likelihood in \eqref{subsample log likelihood}, they proposed an adaptive subsampled MH algorithm. 
\cite{quiroz2018speeding} pointed out that this adaptive sample size determination rule may require large subsample sizes for most MH iterations because the upper bounded $c_r$ may not be sharp enough.  %
In addition, the upper bounded $c_r$ depends on the log-likelihood for the full data, which may require significantly additional computing time in each MH iteration.

The aforementioned work uses uniform subsampling to take subsamples, i.e., all data points have equal probabilities to be included in a subsample, and the focus of the investigations was on the decision rule of subsample sizes. This paper focuses on nonuniform subsampling and shows that it is more efficiency than uniform subsampling, i.e., it produces more accurate approximation with the same subsample size.

\section{MLO Subsampled MH algorithm}
\label{sec:weight-subs-mcmc}
The key to success of the subsampled MH is to approximate the full data log-likelihood $\ell_n(\theta)$ accurately at different values, $\theta$ and $\theta'$, in each iteration using a subsample.  
To improve the approximation efficiency, we propose to use nonuniform subsampling probabilities. 
In this paper, we recommend using sampling with replacement because it has a higher computational efficiency. In addition, if the sampling ratio ($r/n$) is small, then the probability to have duplicates in the subsample is small and thus sampling with replacement has similar estimation efficiency as sampling without replacement.  

Let $\eta_1, ..., \eta_n$ be nonuniform subsampling probabilities such that $\sumn \eta_i=1$. %
For a subsample, $x_1^*, ..., x_r^*$, taken randomly according to $\eta_i$'s with replacement, %
the subsample approximation of $\ell_n(\theta)$ is 
\begin{align}
  \ell_r^*(\theta)
  =\oner\sumr\frac{1}{n\eta_i^*}
  \log\{p(x_i^*|\theta)\}.
\end{align}
Direct calculations show that
\begin{align}
  \Exp^*\{\ell_r^*(\theta)\}
  &=\ell_n(\theta),\label{eq:1}%
  \quad\text{ and }\quad
  \Var^*\{\ell_r^*(\theta)\}
  =\frac{1}{rn^2}\sumn\frac{1}{\eta_i}\log^2\{p(x_i|\theta)\}
  -\frac{1}{r}\ell_n^2(\theta),
\end{align}
where the expectation and variance are taken with respect to the randomness of subsampling only.

Equation~\eqref{eq:1} shows that $\ell_r^*(\theta)$ is an unbiased estimator of $\ell_n(\theta)$. Thus, to better approximate $\ell_n(\theta)$, one can choose $\eta_i$ so that the variance $\Var^*\{\ell_r^*(\theta)\}$ is minimized, that is to find $\bpi^{\opt}=(\eta_1^{\opt}, ..., \eta_n^{\opt})$ such that
\begin{align}\label{eq:6}
  \bpi^{\opt}&=\arg\min_{\bpi}\Var^*\{\ell_r^*(\theta)\}.
\end{align}

Note that
\begin{align*}
  \sumn\frac{1}{\eta_i}\log^2\{p(x_i|\theta)\}%
  &=\sumn\eta_i\times
     \sumn\frac{1}{\eta_i}\log^2\{p(x_i|\theta)\}%
    \ge\bigg[\sumn\big|\log\{p(x_i|\theta)\}\big|\bigg]^2,
\end{align*}
where the second last step is from the Cauchy-Schwarz inequality and the equality holds if and only if when
$\eta_i\propto\big|\log\{p(x_i|\hat\theta)\}\big|$. Thus, the optimal subsampling probabilities that minimize the variance $\Var^*\{\ell_r^*(\theta)\}$ satisfy
\begin{align}\label{eq:6}
  \eta_i^{\opt}%
               =\frac{|\log\{p(x_i|\theta)\}|}
               {\sum_{j=1}^n|\log\{p(x_j|\theta)\}|} \quad i=1, \ldots, n.
\end{align}

Here, $\bpi^{\opt}$ depends on the value of $\theta$, and we use $\bpi^{\opt}(\theta)$ to emphasize this fact when necessary. If in each MH iteration we calculate $\bpi^{\opt}$ for both $\theta$ and $\theta'$, then the computational time is not faster than the full data MH algorithm and there is no computational benefit for using this subsampling plan. 
To address this issue, we propose to calculate $\bpi^{\opt}(\theta)$ at a fixed value of $\theta$ instead of calculating it iteratively. We propose to 
use the maximum likelihood estimator (MLE)
\begin{equation}
  \hat\theta=\arg\max_{\theta}\onen\sumn\log\{p(x_i|\theta)\},
\end{equation}
namely, to use $\bpi^{\opt}(\hat\theta)$ for subsampling in each MH iteration. Heuristically, this is trying to minimize the variance $\Var^*\{\ell_r^*(\theta)\}$ at the value of $\theta$ that are the most likely to occur according to the data. We call this subsampling design the most likely optimal (MLO) subsampling. 

A nice property of the MLO subsampling probability $\bpi^{\opt}$ is that it depends on a fixed value of $\theta$ for a given data set. Thus, we can calculate $\eta_i^{\opt}$ before running subsampled MH algorithm and there is no need to calculate them iteratively. 
In each iteration of the MH algorithm, use the subsample taken according to $\eta_i^{\opt}$ to approximate $\ell_n(\theta)$ and $\ell_n(\theta')$, which are then used to approximate  $\Lambda_n(\theta,\theta')$. We present the procedure in the following algorithm. 

\begin{algorithm}[H]
  \caption{Most likely optimal subsampled Metropolis-Hastings algorithm}
  \label{alg:2}
  \begin{algorithmic}
    \FOR{$k\leftarrow1$ \TO $N$}
    \STATE $\theta \leftarrow \theta_{k-1}$
    \STATE $\theta'\sim q(.\vert\theta)$
    \STATE $u\sim U{(0,1)}$
    \STATE $\psi(u,\theta,\theta')\leftarrow
    \frac{1}{n}\log\Big(u\frac{p(\theta)q(\theta'|\theta)}
    {p(\theta')q(\theta|\theta')}\Big)$
    \STATE $x^*_1,\cdots,x^*_r$ $\; \overset{\bpi^{\opt}}{\sim}X$
    \COMMENT{Subsample with replacement according to $\eta_1^{\opt}, ..., \eta_n^{\opt}$} \\[2mm]
    \STATE $\ell_r^*(\theta)\leftarrow
    \oner\sumr\frac{\log\{p(x_i^*|\theta)\}}{n\eta_i^{\opt *}}$\\[2mm]
    \STATE $\ell_r^*(\theta')\leftarrow
    \oner\sumr\frac{\log\{p(x_i^*|\theta')\}}{n\eta_i^{\opt *}}$\\[2mm]
    \STATE $\Lambda^*(\theta,\theta')\leftarrow
    \ell_r^*(\theta')-\ell_r^*(\theta)$\\[2mm]
    \IF {$\Lambda^*(\theta,\theta')>\psi(u,\theta,\theta')$}
    \STATE $\theta_k\leftarrow\theta'$ \COMMENT{Accept}
    \ELSE
    \STATE $\theta_k\leftarrow\theta$ \COMMENT{Reject}
    \ENDIF
    \STATE Return $\theta_k$, $k=1, ..., N$
    \ENDFOR
    \end{algorithmic}
\end{algorithm}

The performance of Algorithm~\ref{alg:2} critically depends on the quality of $\Lambda^*(\theta,\theta')$ in approximating $\Lambda_n(\theta,\theta')$, which is affected by the subsample size $r$. It is clear that $\Lambda^*(\theta,\theta')$ is unbiased, i.e., $\Exp^*\{\Lambda^*(\theta,\theta')\}=\Lambda_n(\theta,\theta')$. Thus its quality is mainly measured by its variance, which is
\begin{align}
  \Var^*\{\Lambda^*(\theta,\theta')\}
  &=\oner\onen\sumn\frac{\big[\log\{p(x_i|\theta)\}
    -\log\{p(x_i|\theta')\}\big]^2}{|\log\{p(x_i|\hat\theta)\}|}%
    \times\onen\sumn|\log\{p(x_i|\hat\theta)\}|.
\end{align}
Under mild conditions, the Lindeberg-Feller central limit theorem \citep[Section 2.8 of][]{Vaart:98} applies for the conditional distribution of $\Lambda^*(\theta,\theta')$ given $X$, which indicates that, conditional on the full data $X$,
\begin{equation}
  \Lambda^*(\theta,\theta')-\Lambda_n(\theta,\theta')
  \overset{a}{\sim}
  N\big[0,\ \Var^*\{\Lambda^*(\theta,\theta')\}\big],
\end{equation}
where $\overset{a}{\sim}$ means the distribution of the quantity on the left-hand-side is asymptotically the same as the distribution on the right-hand-side. 
This is useful to determine the required subsample size $r$ for a given probability and bound of approximation error. For any given $c_r$ and error probability $\delta_r$, we can approximate the required subsample size by solving
\begin{align}\label{eq:3}
  1-\delta_r&=P\{|\Lambda^*(\theta,\theta')-\Lambda_n(\theta,\theta')|>c_r\}%
      \approx P\big\{\sqrt{\Var^*\{\Lambda^*(\theta,\theta')\}}|Z|>c_r\big\},
\end{align}
where $Z$ is a standard normal random variable. Solving~\eqref{eq:3} gives us the approximated sample size as
\begin{align}\label{eq:4}
  r^a&=\Big(\frac{Z_{\delta/2}}{c_r}\Big)^2
  \onen\sumn\frac{\big[\log\{p(x_i|\theta)\}
    -\log\{p(x_i|\theta')\}\big]^2}{|\log\{p(x_i|\hat\theta)\}|}%
                                                                     \times\onen\sumn|\log\{p(x_i|\hat\theta)\}|.%
\end{align}

Of course, direct use of equation~\eqref{eq:4} is not computationally appealing as it requires to evaluate the likelihood on the full data. We can use a pilot subsample to estimate $r^a$ and then decide if we need additional data to achieve the pre-specified level of precision. An unbiased estimator of $r^a$ based on a subsample $x_1^*, ..., x_r^*$ is 
\begin{equation}\label{eq:8}
  r^{a*}=\Big(\frac{Z_{\delta/2}}{c_r}\Big)^2
  \frac{1}{rn^2}\sumr\frac{\big[\log\{p(x_i^*|\theta)\}
    -\log\{p(x_i^*|\theta')\}\big]^2}{(\eta_i^{\opt *})^2}.
\end{equation}
Based on \eqref{eq:8}, we propose an adaptive version of the most likely optimal subsampled MH algorithm presented below. 
\begin{algorithm}[H]

  \caption{Adaptive most likely optimal subsampled MH algorithm}
  \label{alg:3}
  \begin{algorithmic}
    \FOR{$k\leftarrow1$ \TO $N$}
    \STATE $\theta \leftarrow \theta_{k-1}$
    \STATE $\theta'\sim q(.\vert\theta)$
    \STATE $u\sim U{(0,1)}$
    \STATE $\psi(u,\theta,\theta')\leftarrow
    \frac{1}{n}\log\Big(u\frac{p(\theta)q(\theta'|\theta)}
    {p(\theta')q(\theta|\theta')}\Big)$
    \STATE $x^*_1,\cdots,x^*_r$ $\; \overset{\bpi^{\opt}}{\sim}X$
    \COMMENT{Subsample with replacement according to $\eta_1^{\opt}, ..., \eta_n^{\opt}$} \\[2mm]
    \STATE $\ell_r^*(\theta)\leftarrow
    \oner\sumr\frac{\log\{p(x_i^*|\theta)\}}{n\eta_i^{\opt *}}$\\[2mm]
    \STATE $\ell_r^*(\theta')\leftarrow
    \oner\sumr\frac{\log\{p(x_i^*|\theta')\}}{n\eta_i^{\opt *}}$\\[2mm]
    \STATE $\Lambda^*(\theta,\theta')\leftarrow
    \ell_r^*(\theta')-\ell_r^*(\theta)$\\[2mm]
    \STATE  $c_r=|\Lambda^*(\theta,\theta')-\psi(u,\theta,\theta')|/2$
    \STATE  $r^{a*}\leftarrow\Big(\frac{Z_{\delta/2}}{c_r}\Big)^2
  \frac{1}{rn^2}\sumr\frac{\big[\log\{p(x_i^*|\theta)\}
    -\log\{p(x_i^*|\theta')\}\big]^2}{(\eta_i^{\opt *})^2}$\\[2mm]
    \IF {$r<r^{a*}\land r_{\text{max}}$}
    \STATE $x^*_{r+1},\cdots,x^*_{r^{a*}}$ $\; \overset{\bpi^{\opt}}{\sim}X$
    \COMMENT{Take additional subsample} \\[2mm]
    \STATE $\ell_r^*(\theta)\leftarrow
    \frac{1}{r^{a*}}\sum_{i=r+1}^{r^{a*}}\frac{\log\{p(x_i^*|\theta)\}}{n\eta_i^{\opt *}} + \frac{r}{r^{a*}}\ell_r^*(\theta)$\\[2mm]
    \STATE $\ell_r^*(\theta')\leftarrow
    \frac{1}{r^{a*}}\sum_{i=r+1}^{r^{a*}}\frac{\log\{p(x_i^*|\theta')\}}{n\eta_i^{\opt *}} + \frac{r}{r^{a*}}\ell_r^*(\theta')$\\[2mm]
    \STATE $\Lambda^*(\theta,\theta')\leftarrow
    \ell_r^*(\theta')-\ell_r^*(\theta)$\\[2mm]
    \ENDIF
    \IF {$\Lambda^*(\theta,\theta')>\psi(u,\theta,\theta')$}
    \STATE $\theta_k\leftarrow\theta'$ \COMMENT{Accept}
    \ELSE
    \STATE $\theta_k\leftarrow\theta$ \COMMENT{Reject}
    \ENDIF
    \STATE Return $\theta_k$, $k=1, ..., N$
    \ENDFOR
    \end{algorithmic}
\end{algorithm}

\section{Simulation}
\label{sec:simulation}
In this section, we use numerical experiments to evaluate the performance of the proposed MLO subsampled MH algorithm, and compare it with the uniform subsampled MH algorithm.

We repeat the simulation for $B=100$ times and calculate the empirical bias, standard deviation, and mean squared error as
\begin{align*}
  \begin{split}
    \text{Bias} &= \frac{1}{B}\sum_{b = 1}^B\hat{\theta}_b-\theta,
    \quad
    \text{SD} = \frac{1}{B-1}\sum_{b = 1}^B (\hat{\theta}_b -
    \bar{\hat{\theta}}),
    \quad\text{ and}\quad
    \text{MSE} = \frac{1}{B}\sum_{b=1}^B(\hat{\theta}_b - \theta)^2,
  \end{split}
\end{align*}
respectively, where $\hat{\theta}_b$ is the estimate in the $b$th repetition of the simulation, $\theta$ is the true parameter, and $\bar{\hat{\theta}}$ is the mean of the estimates from the $B$ repetitions of the simulation. 

\subsection{Example \uppercase\expandafter{\romannumeral1}: Gaussian distribution}

We generate 1000 observations from $\text{N}(\mu,1)$ with true $\mu=1$, and choose the prior as $\mu\sim \text{N}(0,3^2)$ where $3^2$ is the variance of the prior. We run $N=$3,000 iterations of subsampled MH algorithms and throw away the samples from the first 1,000 iterations as burn-in and use the sample mean for the rest of the sample to estimate $\mu$. %

In addition to using the MLE of $\mu$ to calculate the subsampling probability $\bm{\eta}$, we investigate the effect of using different values of $\mu$. We first considered $\mu=1$, which is the true value of the parameter in generating data sets. In addition, we consider some other values of $\mu$: $\mu=$ -10, -5, -2, 1, 2, 5 and 10. 

Table \ref{table:mean_comp} presents the results of empirical mean squared error, in which $\bm{\eta}_{\hat{\mu}}^*$ and $\bm{\eta}_{\hat{\mu}}^t$ are the subsampling probabilities calculated using the MLE from the data and using the true parameter, respectively. It is seen that $\bm{\eta}_{\hat{\mu}}^*$ and $\bm{\eta}_{\mu=1}^t$ (based on the true value of $\mu$ in generating the data) have very similar performances, and they both outperform other choices of $\bm{\eta}$. Note that the true value $\mu=1$ is always unknown in practice, but using the practical $\bm{\eta}_{\hat{\mu}}^*$ produces comparable results. Compared with other choices of $\mu$, the advantage of $\bm{\eta}_{\hat{\mu}}^*$ is more significant for smaller values of the subsample size $r$. 
\begin{table}
  \center
  \caption{Empirical MSE ($\times 10^3$) with different weights for estimating the mean parameter in a Gaussian distribution}
  \label{table:mean_comp}
  \begin{tabular}{ccccccccc}
    \hline	&$\bm{\eta}_{\hat{\mu}}^*$&$\bm{\eta}_{\mu=1}^t$&$\bm{\eta}_{\mu=-10}$&$\bm{\eta}_{\mu=-5}$&$\bm{\eta}_{\mu=-2}$&$\bm{\eta}_{\mu=2}$&$\bm{\eta}_{\mu=5}$&$\bm{\eta}_{\mu=10}$\\
    \hline
    $r=10$&1.53&1.51&1.96&3.22&10.1&3.06&6.43&2.14\\
    $r=20$&1.46&1.34&1.41&2.06&3.86&1.76&2.61&1.56\\
    $r=50$&1.23&1.25&1.33&1.33&1.78&1.31&1.46&1.36\\
    \hline	
  \end{tabular}	
\end{table}

Similar to the case of the mean parameter, we implemented the nonuniform subsampling method in estimating the precision parameter (the inverse of the variance) of the normal distribution. We generate 1000 observations from $\text{N}(0,\tau)$ with $\tau=1$, and choose the prior as $\tau \sim \text{Gamma}(0.01,0.01)$. Here $\tau$ is the precision parameter. We run 3,000 iterations of subsampled MH algorithms and throw away the samples from the first 1,000 iterations as burn-in. We consider using different values of $\tau$ to calculate the subsampling probability, with the choices of the true parameter $\tau=1$, and $\tau=0.1, 0.2, 0.5, 2, 5,$ and $10$.
Table \ref{table:variance_comp} presents the results on empirical MSE. Again, $\bm{\eta}_{\hat{\tau}}^*$ performs similarly to $\bm{\eta}_{\tau=1}^t$ and it is better than other choices of $\tau$. 
\begin{table}
  \center
  \caption{Empirical MSE ($\times 10^3$) with different weights for estimating the precision parameter in a Gaussian distribution.}
  \label{table:variance_comp}
  \begin{tabular}{ccccccccc}
    \hline
    &$\bm{\eta}_{\hat{\tau}}^*$&$\bm{\eta}_{\tau=1}^t$&$\bm{\eta}_{\tau=0.1}$&$\bm{\eta}_{\tau=0.2}$&$\bm{\eta}_{\tau=0.5}$&$\bm{\eta}_{\tau=2}$&$\bm{\eta}_{\tau=5}$&$\bm{\eta}_{\tau=10}$\\
    \hline
 $r=10$ & 2.16 & 2.40 & 5.19 & 3.97  & 2.86 & 2.69 & 48.7 & 100  \\
 $r=20$ & 2.16 & 2.02 & 3.18 & 2.40  & 2.22 & 2.42 & 18.5 & 62.2 \\
 $r=50$ & 1.93 & 1.97 & 2.23 & 2.24  & 2.02 & 1.94 & 4.64 & 23.9 \\
    \hline

  \end{tabular}
\end{table}

We also compare the performance of our method with the uniform subsampled MH algorithm. Table~\ref{precision} shows the empirical results, where the proposed method are significantly better than the uniform subsampling method. In this case, the bias is not negligible for the uniform subsampling method, while the proposed method has very small bias.  

\begin{table}
  \caption{Empirical bias, standard deviation, and root mean square errors for estimating the precision of a Gaussian distribution. All number are multiplied by $\times 10^3$ for better presentation.}
  \center
  \label{precision}
  \center
  \begin{tabular}{cccccccc}
    \hline
    & \multicolumn{3}{c}{MLO} && \multicolumn{3}{c}{Uniform subsampling}\\
    \cline{2-4} \cline{6-8}
    Subsample Size &Bias& SD &$\sqrt{\text{MSE}}$&&Bias& SD &$\sqrt{\text{MSE}}$\\
    \hline
    $r=10$    & -2.86 & 48.2 & 48.1 &  & 63.0  & 51.8 & 81.4 \\
    $r=20$    & -3.03 & 45.2  & 45.2 &  & 29.2 &50.8 & 58.4 \\
    $r=50$    & -1.57 & 45.5 & 45.5 &  & 9.91 & 44.0 & 44.8 \\
    \hline
  \end{tabular}
\end{table}

\begin{figure}
  \subfigure[$r=10$]{
    \includegraphics[width=0.3\textwidth]{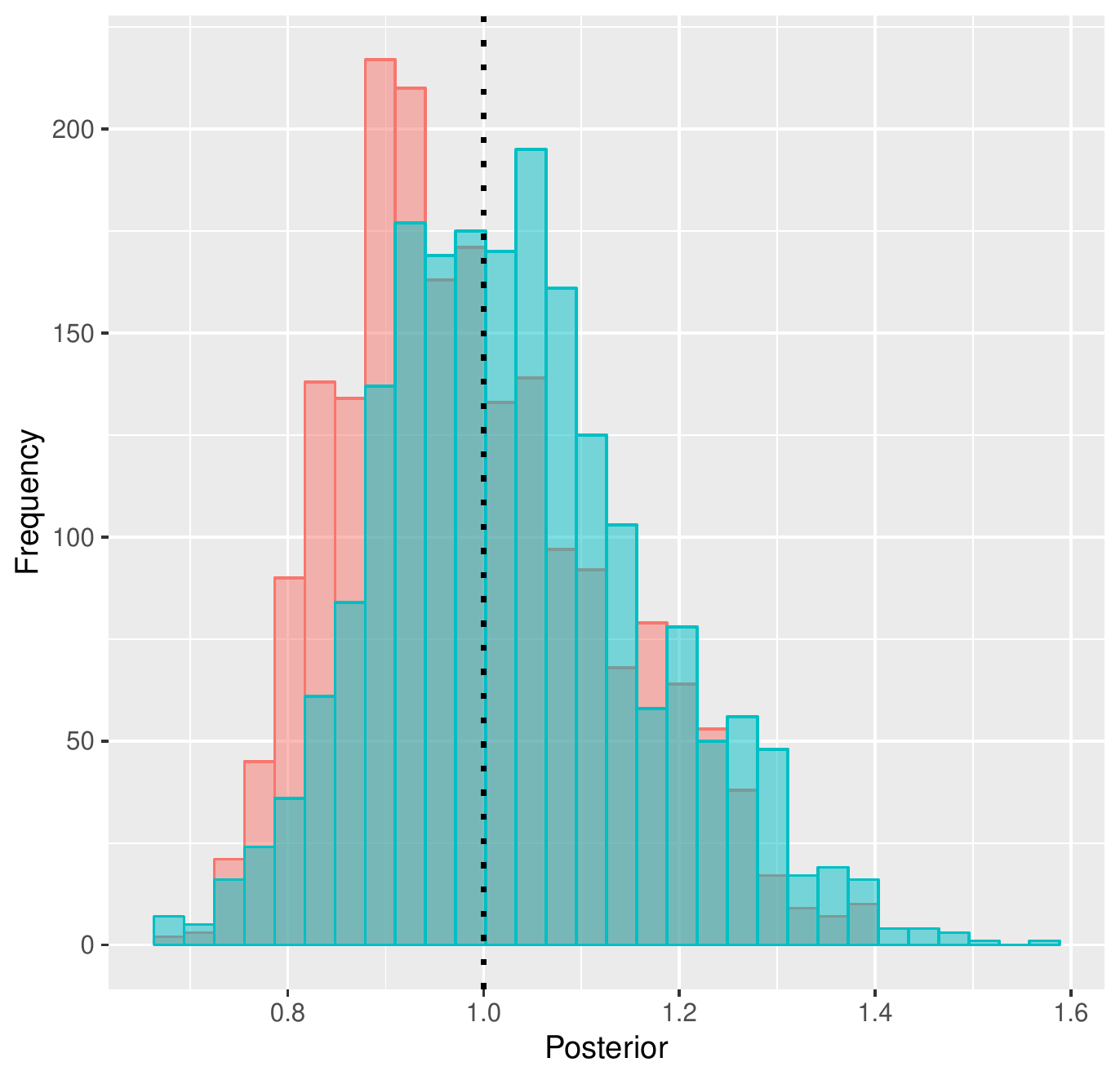}
    }
     \subfigure[$r=20$]{
    \includegraphics[width=0.3\textwidth]{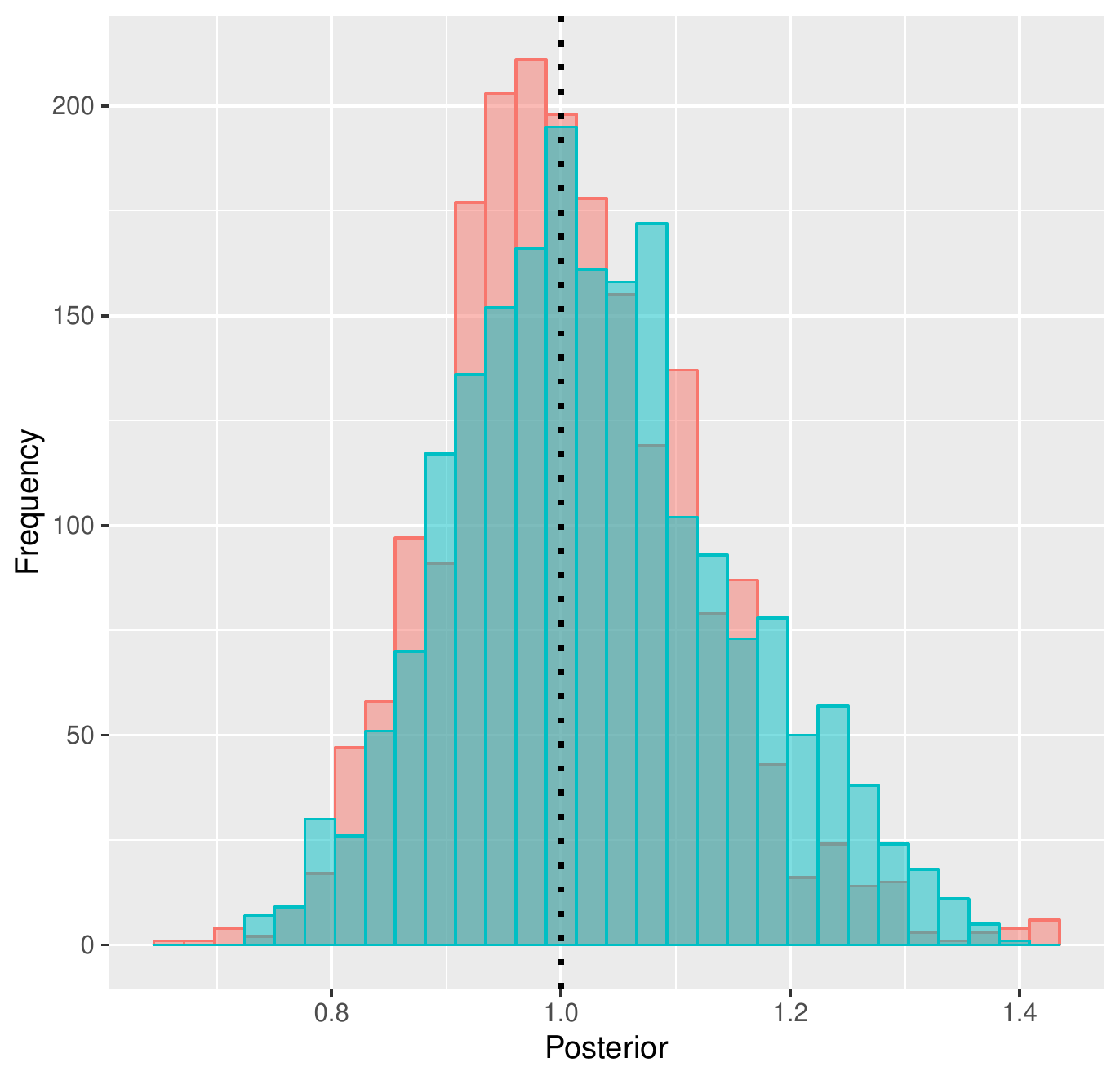}
    }
     \subfigure[$r=50$]{
    \includegraphics[width=0.3\textwidth]{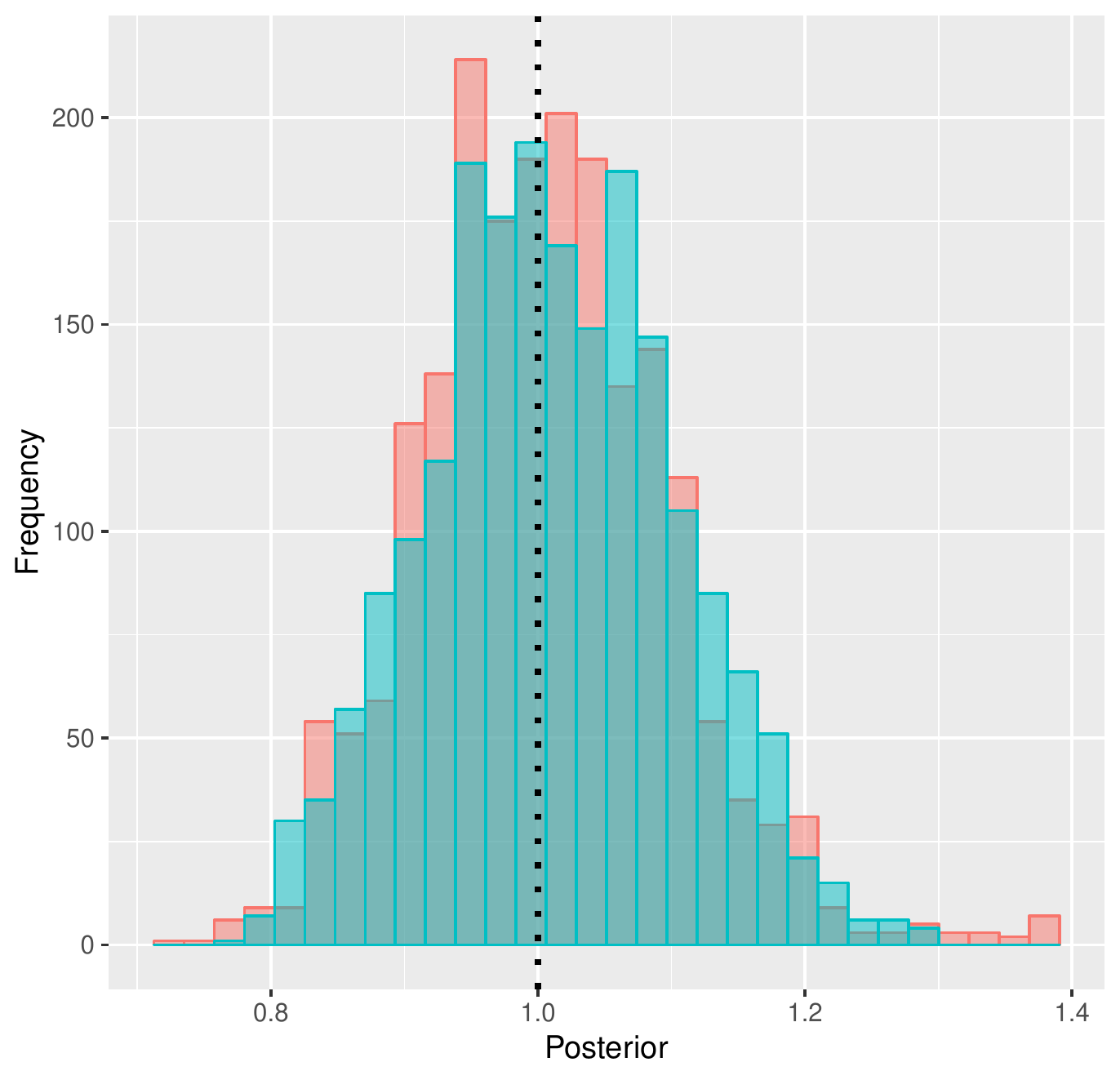}
    }
  \caption{Histogram of posterior samples from subsampled MH algorithms with $r=10, 20,$ and $50$, respectively. The red color is for the MLO subsamples and the green color is for the uniform subsamples.}
\label{fig:posterior_hist}
\end{figure}

To have a closer look at the difference between posterior samples from the MLO subsampled MH algorithm and those from the uniform subsampling MH algorithm, we plot histograms using 2,000 posterior samples from each algorithm in Figure \ref{fig:posterior_hist}. 
We see that posterior samples from the MLO subsampled MH algorithm have smaller variances than those from the uniform subsampled MH algorithm. The mean of the MLO subsampled posterior samples is also closer to the true posterior mean (the vertical dotted line) compared with that of the uniform subsampled posterior samples.

\subsection{Example \uppercase\expandafter{\romannumeral2}: logistic regression}
In this example, we compare our MLO subsampled MH algorithm with the uniform subsampled MH algorithm in the context of logistic regression. 
In each repetition of the simulation, we generate $n=$100,000 data points from a logistic regression model. Specifically, for each repetition, data points $x_i=(y_i,z_i)$, $i=1, ..., 100,000$, are generated in this way: generate $z_i=(z_{1i}, z_{2i})$ so that $z_{1i}$ and $z_{2i}$ are independent and identically distributed from the standard normal distribution; generate $y_i$ independently from Bernoulli distributions $\text{Bernoulli}(\gamma_i)$, where $\gamma_i=\frac{\text{exp}(z_{1i}\theta_1+z_{2i}\theta_2)}{1+\text{exp}(z_{1i}\theta_1+z_{2i}\theta_2)}$, and the true parameters are $\theta_1=1$ and $\theta_2=0.5$. 

For the two parameters $\theta_1$ and $\theta_2$, we use posterior means to estimate them, which are Bayes estimators under the squared loss.  The prior distributions of $\theta_1$ and $\theta_2$ are both assumed to be Gaussian distribution with mean 0 and variance 10.  %
We use subsampled MH algorithms to draw samples from the posterior to approximate the posterior means. For the proposal distribution $q(\theta'|\theta)$, we set it to be $N(\theta,1)$ %
which corresponds to the random walk MH algorithm \cite{haario1999adaptive}. For each repetition of the simulation, we run 30,000 MH iterations and throw away the first 10,000 samples as burn-in and store one sample for every 20 MH iterations in order to reduce autocorrelation. We repeat the simulation for 100 times, and calculate the empirical biases and standard errors for the posterior means. Results are reported in Table~\ref{tab:1}. For the adaptive algorithm in Algorithm~\ref{alg:3}, the initial subsample size is $r=100$ and the upper limit of subsample size $r_{\text{max}}=5,000$. The mean subsample percentage of the adaptive algorithm in Algorithm~\ref{alg:3} is about 1.68 \%, and the median percentage is about 0.96 \%.
\begin{table}
\center
\label{simu_logistic2}
\caption{Empirical bias $\times10^3$ and empirical standard deviation $\times10^3$ for estimating parameters in logistic regression. %
}\label{tab:1}
\begin{tabular}{rcccccc}
  \hline
  $r/n$ & method &\multicolumn{2}{c}{$\theta_1$}&\multicolumn{2}{c}{$\theta_2$}\\
  \hline
                 &            & Bias & SD   & Bias & SD   \\\hline
 0.001           & uniform    & 60.6 & 11.9 & 30.1 & 12.6 \\
                 & MLO & 15.4 & 13.4 & 6.58 & 12.1 \\
 0.002           & uniform    & 39.5 & 11.8 & 19.8 & 9.96 \\
                 & MLO & 17.1 & 11.7 & 8.30 & 9.99 \\
 0.005           & uniform    & 19.5 & 10.2 & 9.93 & 8.11 \\
                 & MLO & 9.56 & 10.1 & 5.21 & 8.91 \\
 0.01            & uniform    & 10.1 & 11.2 & 5.98 & 7.91 \\
                 & MLO & 5.85 & 8.99 & 3.74 & 8.11 \\
 $\approx0.0168$ & Adaptive   & 2.57 & 10.4 & 1.78 & 7.69 \\
  \hline
\end{tabular}
\end{table}

        From Table~\ref{tab:1}, with the same subsampling ratio, our MLO subsampled algorithm is better than the uniform subsampled algorithm for both $\theta_1$ and $\theta_2$. Estimators from the the two different subsampled algorithms may have similar standard errors, but estimators from MLO subsampled MH algorithm have much smaller biases.

\begin{figure}
	\center
	\includegraphics[width= 3.5 in]{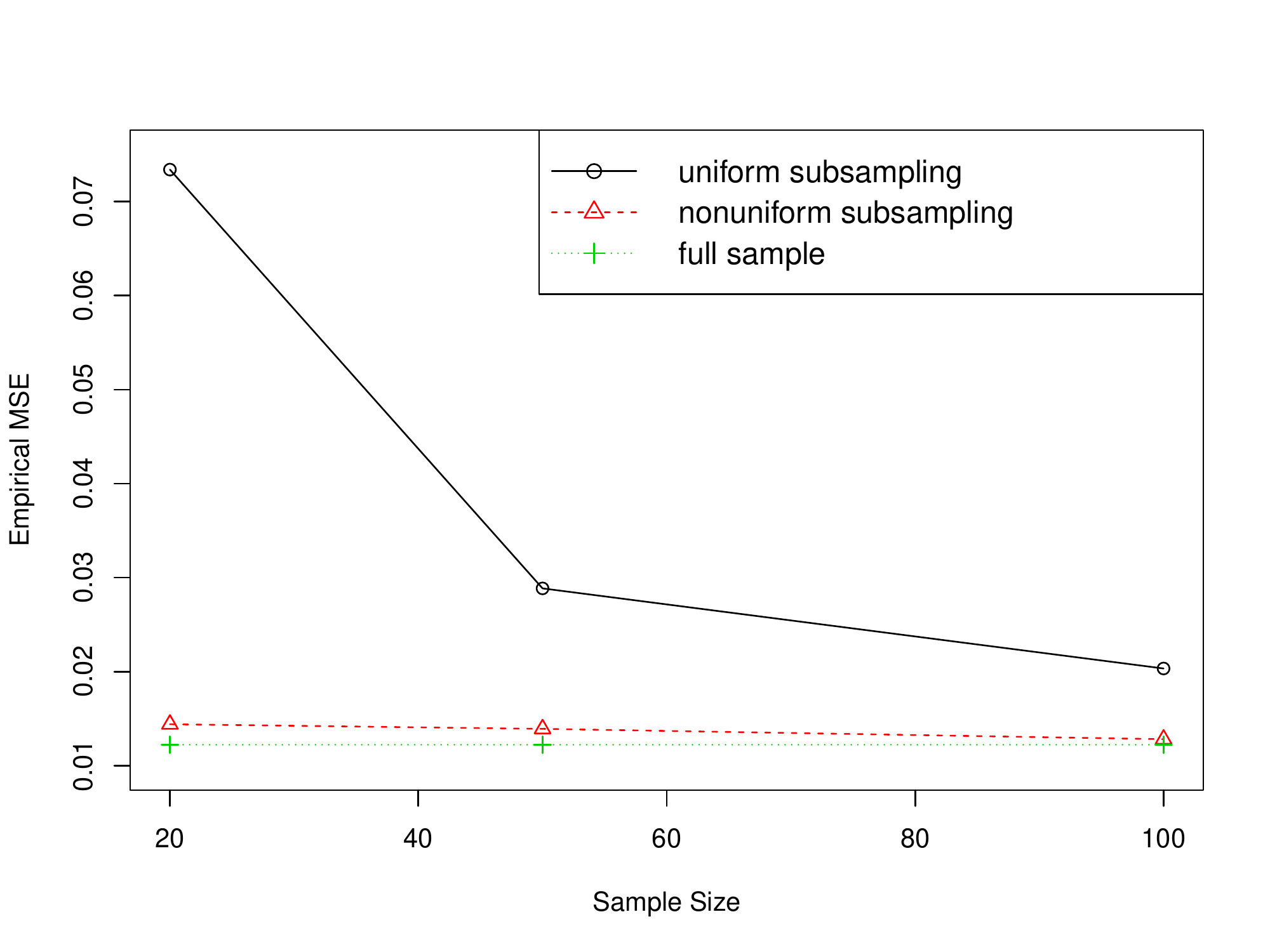}
	\caption{Empirical MSE versus subsample size for logistic regression.}
	\label{fig1:mse}
\end{figure}

In order to compare the estimation efficiency of the subsampled algorithms with that of full sample MH algorithm, we reduce the full data sample size to $n=1000$ so that the full sample MH algorithm are computationally tractable. Except this change, other simulation configurations are the same as the previous case. Figure \ref{fig1:mse} shows the sum of the empirical mean squared error for $\theta_1$ and $\theta_2$ with different subsampling methods. 
We see that even with subsample size $r=20$, i.e., using 2\% of the full sample, our MLO subsampled MH algorithm has similar estimation efficiency to the full sample MH algorithm. On the other hand, the uniform subsampled MH algorithm requires a much larger subsample size to achieve the same level of estimation efficiency.

\section{Covertype Data}
\label{sec:covertype-data}
We apply the MLO subsampled MH algorithm to the UCI Covertype data set. Like \cite{collobert2002parallel}, we convert the multiple classification problem into a binary classification problem by focusing on predicting one class of the responses. We use the 10 continuous covariates to model the probability that the response belongs to this class through a logistic regression model with an intercept. %
The total number of data points in this data set is $n=$581,012.  We run 500,000 MH iterations, and throw the first 100,000 samples as burn-in and keep one sample for every 20 iterations after the burn-in. For Algorithm~\ref{alg:3}, we choose the initial subsample size as $r=1,000$ and $r_{\text{max}}=20,000$, and the average subsample size is around 14,000. We report posterior means, posterior standard deviations and $95\%$ highest posterior density (HPD) intervals \citep{chen1999monte} for parameters in logistic regression in following Table \ref{covertype_result_ps}. 
We also obtain posterior estimates form the uniform subsampled MH algorithm for comparisons, and we set $r=15,000$ for this approach.  %
\begin{table}
  \caption{Results for the covertype data set}
  \label{covertype_result_ps}
  \center
  \begin{tabular}{cccc}
    \hline
    & \multicolumn{3}{c}{MLO subsampling} \\
    \cline{2-4} 
    Parameter     & Posterior Mean & SD    & $95\%$ HPD interval \\
    \hline
    $\theta_0$    & -2.277         & 1.020 & (-4.354, -0.477)    \\
    $\theta_1$    & -0.525         & 0.020 & (-0.564, -0.487)    \\
    $\theta_2$    & -0.069         & 0.022 & (-0.113, -0.027)    \\
    $\theta_3$    & 0.104          & 0.041 & (0.026, 0.181)      \\
    $\theta_4$    & 0.262          & 0.024 & (0.216, 0.309)      \\
    $\theta_5$    & -0.051         & 0.022 & (-0.097, -0.007)    \\
    $\theta_6$    & 0.126          & 0.019 & (0.089, 0.161)      \\
    $\theta_7$    & 0.572          & 0.160 & (0.304, 0.920)      \\
    $\theta_8$    & 0.018          & 0.094 & (-0.192, 0.181)     \\
    $\theta_9$    & 0.547          & 0.186 & (0.232, 0.952)      \\
    $\theta_{10}$ & 0.285          & 0.019 & (0.246, 0.323)      \\
    \hline
    & \multicolumn{3}{c}{uniform subsampling}\\
    \cline{2-4}
    Parameter     & Posterior Mean & SD    & $95\%$ HPD interval \\
    \hline
    $\theta_0$    & -3.43          & 1.129 & (-6.161, -1.478)    \\
    $\theta_1$    & -0.525         & 0.025 & (-0.576, -0.476)    \\
    $\theta_2$    & -0.067         & 0.030 & (-0.129, -0.011)    \\
    $\theta_3$    & 0.144          & 0.046 & (0.056, 0.249)      \\
    $\theta_4$    & 0.256          & 0.027 & (0.205, 0.309)      \\
    $\theta_5$    & -0.044         & 0.026 & (-0.095, 0.004)     \\
    $\theta_6$    & 0.124          & 0.017 & (0.091, 0.156)      \\
    $\theta_7$    & 0.769          & 0.157 & (0.471, 1.136)      \\
    $\theta_8$    & -0.101         & 0.090 & (-0.310, 0.079)     \\
    $\theta_9$    & 0.779          & 0.183 & (0.472, 1.242)      \\
    $\theta_{10}$ & 0.290          & 0.023 & (0.245, 0.336)      \\
    \hline
  \end{tabular}
\end{table}

In Table~\ref{covertype_result_ps}, the uniform subsampled MH algorithm results in larger posterior standard deviations than the adaptive MLO subsampled MH algorithm with similar subsample sizes. From the uniform subsampled MH algorithm, $\theta_5$ is not significant, which is not consistent with the result from the full data MLE confidence interval. We also observe that the MLO subsampled MH algorithm converged faster to the posterior distribution than uniform subsampled MH algorithm for this data set. Thus, we need less number of iterations of the MH algorithm with MLO subsampling for the same level of Monte Carlo error.

\section{Conclusion}
\label{sec:conclusion}
In this paper, we have developed subsampled MH algorithms with most like optimal subsampling probabilities in approximating the full data log-likelihood. We have also provide a rule to determine the required subsample size in each MH iteration adaptively. In experiments based on both simulated and real data sets, our MLO subsampled MH algorithms have outperformed the uniform subsampled MH algorithm.

We conclude this paper by pointing to some directions
for future research. 
First, model selection has always been an important topic in statistical analysis, but this topic has not been investigated in the context of data-dependent subsampling. Thus developing Bayesian model selection criteria \cite{geng2019subsampled} for subsampling algorithm is a desirable future research topic. Second, in this paper we have only considered sampling the data to approximate the likelihood. A subsampling strategy to sample from the posterior distribution warrants further research. Third, we assume that the full data are available all at once in this paper. Developing subsampled Bayesian estimation procedures in an online learning setting is an interesting task with significantly practical value.

\bibliography{ref}

\bibliographystyle{chicago}

\end{document}